\documentclass[12pt,preprint]{aastex}

\shorttitle{Velocity Profiles in the Solar Corona}
\shortauthors{Qu\'emerais et al.}

\usepackage{graphicx}

\def\rso{$R_{\odot}$}

\begin{document}

%
%

\title{Velocity Profiles in the Solar Corona from Multi-Instrument Observations}


%
%

\author{E. Qu\'emerais, R. Lallement, D. Koutroumpa} 
\affil{Service d'A\'eronomie, Verri\`eres le Buisson, France}

\and

\author{P. Lamy}
\affil{Laboratoire d'Astrophysique de Marseille, Marseille, France}

%
%
%

\begin{abstract}

We present a method to derive outflow velocities in the solar corona using different data sets
including solar wind mass flux coming from the SWAN/SOHO instrument, electron density values
from  LASCO-C2 and interplanetary solar wind velocities derived from ground-based 
Interplanetary Scintillation Observations (IPS). 

In a first step, we combine the LASCO electron densities at 6 \rso\ and the IPS velocities, and compare 
the product to the SWAN mass fluxes.  It is found that this product 
represents the actual mass flux at 6 \rso\ for the fast wind, but not for the slow wind.
In regions dominated by the slow wind, the fluxes derived from SWAN are systematically smaller. 
This is interpreted as a proof that the fast solar wind has reached its terminal velocity
 at $\sim$ 6 \rso\ and expands with constant velocity beyond this distance. On the contrary, the slow 
solar  wind has reached only half of its terminal  value and is thus accelerated further out.

In a second step, we combine the LASCO-C2 density profiles and the SWAN flux data to derive velocity 
profiles in the corona between 2.5 and 6 \rso\ . Such profiles can be used to test
models of the acceleration mechanism of the fast solar wind.

\end{abstract}


\section{Introduction}

There is no consensus on the mechanisms of energy deposition and solar wind acceleration in the 
low corona, especially in the fast wind, although major results  have been obtained with the SOHO 
instruments. In particular, UV spectroscopy has led to the discovery of high ion temperatures 
and significantly lower electronic temperatures in coronal holes, as well as strong ion temperature 
anisotropies (e.g., Kohl et al., 2006; David et al., 1998). A number of models have been developed to 
reproduce these properties, with damping of high frequency waves and ion cyclotron heating as favoured 
mechanisms.  Because one of the strong difficulties in modeling the solar wind expansion is linked to 
 the heat flux in regions which are not collisional enough for classical transport 
theory to be valid, kinetic models have also been developed. These models  are able to reproduce the 
acceleration by velocity filtration (e.g., Zouganelis et al., 2005), although they do not reproduce the 
published temperature anisotropies. Since all different classes of models predict different velocity 
profiles, accurate measurements of the solar wind flow speed in the acceleration region are required to 
decide which models are correct.

One of the methods to obtain the flow speed profile in the acceleration region  relies on the measured 
coronal density profile as well as the expanding geometry. The velocity profile is simply obtained from 
the application of the continuity equation, assuming a value for the particle flux coming from the observed region. 
The  flux is inferred from in situ or remote sensing measurements. Munro and Jackson (1977) pioneered 
this method using Skylab white-light coronographic images of polar coronal holes. Assuming that the solar
flux is roughly constant at all latitudes, they used its average ecliptic value. They concluded that there 
is a substantial energy addition in the low corona, and that the flow accelerates to supersonic speed 
within 2-3\rso\ .  

Later on, the analysis of solar backscattered Lyman-alpha radiation led to the derivation of a 
significantly lower mass flux in the polar regions (Kumar and Broadfoot, 1979; Lallement et al., 1985a).
It was  shown that the requirement for a non-thermal mechanism is strongly dependent on the density 
profile and on the out-of-ecliptic mass flux (Lallement et al., 1985b). 
However, our knowledge of the solar wind conditions outside of the ecliptic plane is very limited as
most of the in situ studies of the solar wind are based on data obtained in the ecliptic plane. 
Because of the small angle (7 degrees) between the ecliptic plane and the solar equatorial 
plane, this means that most of our knowledge on the solar wind concerns the wind which originates 
close to the solar equator. 


In this article, we present a new approach to the determination of velocity profiles in the solar corona.
The steps are the following,
\begin{itemize}
\item
The velocity at large distance from the Sun is obtained from Interplanetary Scintillation data (Kojima et al., 1998).
\item
Using the SWAN interplanetary Lyman $\alpha$ data, we can derive the ionization rate of hydrogen at 1 AU. 
After removal of the contributions from photo-ionization and electron impact ionization, we obtain the ionization
rate due to charge exchange with the solar wind protons.
\item
Using the value of the charge exchange cross section, which depends on solar wind velocity, we compute the solar wind
flux.
\item
Assuming radial expansion, we derive the solar wind velocity profile from the values of coronal density and solar wind mass
flux. The assumption of radial expansion is not necessary. We could introduce a correction term for non-radial expansion.
\item
The previous relation is then applied to the density profiles obtained from the LASCO-C2 polarized white light images, between
2.5\rso\ and 6\rso\ .
\end{itemize}

%

In what follows, we concentrate on a test period from 1996 to 1997 corresponding to a minimum in the solar 
activity cycle. We briely present the results of Qu\'emerais et al. (2006) for this period. We then show how 
mean values of density and velocity can be derived. In the last paragraph, we show how this information can 
be combined with density profiles in the solar corona to give information on velocity profiles. 

In our conclusion, we discuss the velocity profiles found in the corona and how they relate to fast wind or 
slow wind  conditions. These results are relevant to studies of the mechanism of the solar wind acceleration.

\section{Method of Analysis}

The Lyman $\alpha$ background is due to the backscatter of solar Lyman $\alpha$ photons by hydrogen atoms
present in the interplanetary medium. This UV emission was first identified by instruments on board the OGO-8 
spacecraft (Bertaux and Blamont, 1971; Thomas and Krassa, 1971). The hydrogen atoms of the local insterstellar 
cloud can get very close to the Sun because of the motion of the solar system relative to the local cloud. 
In the vicinity of the sun, the hydrogen atoms are ionized by charge exchange with solar wind protons or 
photo-ionized by solar EUV radiation or, with a lesser probability, ionized by collision
with hot solar wind electrons. 

In the inner heliosphere the spatial pattern of the full-sky Lyman $\alpha$  background is carved by the 
ionizing fluxes from the Sun (Jozelyn and Holzer, 1975; Kumar and Broadfoot, 1979). 
Once a hydrogen atom loses its electron, it cannot backscatter Lyman $\alpha$ photons anymore. This creates a hole 
in the sky pattern. Areas of the sky where ionization is more effective appear dimmer at Lyman $\alpha$ because 
there are less hydrogen atoms which can backscatter solar Lyman $\alpha$ photons.  The most effective ionization 
process, roughly 4 times out of 5, is charge exchange with solar wind protons. Even in that case, the hydrogen atom 
is lost because the hydrogen atom which captured the electron has the velocity of the solar wind proton, i.e. 
at least a few hundred km/s. The illuminating solar Lyman $\alpha$ line is roughly 1 Angstr\"om wide. 
In terms of Doppler shift, this corresponds to 150 km/s. This means that a hydrogen atom with a velocity 
larger than 150 km/s is blind to solar Lyman $\alpha$ photons and this results in a loss of hydrogen atoms able 
to backscatter the solar atoms exactly as in the case of photo-ionization.

Qu\'emerais et al. (2006) published an analysis of the full-sky background maps obtained with SWAN/SOHO 
between 1996 and 2005. The
maps were used to derive the hydrogen ionization rate at 1 AU at all heliographic latitudes. We will use these 
values to derive the solar mass flux and its latitudinal dependence.  The ionization rate of hydrogen is the sum
of three terms, the photo-ionization rate, the charge exchange ionization rate, and the electron impact ionization 
rate. Photo-ionization and charge exchange have a radial dependence like the inverse of the square of the distance 
to the sun. The photon flux and the solar particle flux have approximately the same radial variation. 
The electron impact ionization rate also varies with the solar wind electron temperature which is not constant with 
distance to the sun so its total radial dependence is not like the inverse of the square of the distance to the sun.
Using the solar wind electron distribution values obtained by the WIND/waves experiment (Issautier et al., 2004)
we have computed the absolute value of the electron impact ionization rate at 1 AU between 1994 and 2001. 
We have found values between 10$^{-8}$ s$^{-1}$ at solar minimum and a few  10$^{-8}$ s$^{-1}$ at solar maximum. 
The total ionization rate found by Qu\'emerais et al. (2006) varies between 6 10$^{-7}$ s$^{-1}$ and 
1.2 10$^{-6}$ s$^{-1}$, i.e. almost two orders of magnitude larger.
Then, because electron impact ionization is small compared to the two other sources of ionization, we will
neglect it and assume that the total ionization rate of hydrogen varies like the inverse of the square of the 
distance to the sun. 

The total ionization rate $\beta_{tot}$ at 1 AU can be written as

\begin{equation} \label{basicbeta}
\beta_{tot} = \beta_{exc} + \beta_{phot} + \beta_{eii} = \sigma_{exc}(v_{sw}) \cdot v_{sw} \cdot N _{sw} + 
\beta_{phot} + \beta_{eii}
\end{equation}

In the previous equation, we neglect the velocity of the hydrogen atom before the solar wind speed $ v_{sw}$. 
The charge exchange cross section variation with velocity $\sigma_{exc}(v_{sw})$ is given by 

\begin{equation} \label{basicbeta2}
\sigma_{exc}(v_{sw}) \cdot v_{sw}  = \frac{\beta_{tot} - \left( \beta_{phot} + \beta_{eii} \right)}{N _{sw}}
\end{equation}

Equation \ref{basicbeta2} can be used to derive the solar wind mass flux from the knowledge of $\beta_{exc}$ and
the density of the solar wind. In the range 0-1200 km/s the product $\sigma_{exc}(v_{sw}) \cdot v_{sw}$ is a 
monotonic function of the velocity $v_{sw}$, and can easily be inverted. The charge exchange ionization rate 
is obtained from the SWAN data analysis once a value for photo-ionization and electron impact ionization has 
been removed.


The velocity profile $V(R)$ is simply derived using the flux conservation which can be written as follows, 
where $R$ is the distance from the Sun center (in \rso\ ), and $R_o$ is 1 AU,
\begin{equation}
 N(R) \cdot V(R) \cdot R^2 =  N_o \cdot V_o \cdot R_o^2 
\end{equation}

In that case the the velocity profile is given by

\begin{equation}
V(R) = \frac{\beta_{exc}}{\sigma_{exc}(V_o)} ~ \frac{R_o^2}{R^2} ~ \frac{1}{N(R)}
\end{equation}

This formalism will be used in the following sections.

\section{Data Sets}

\subsection{Solar Wind Mass Flux: SWAN data}

The Solar Wind ANisotropy (SWAN) instrument on-board the SOHO spacecraft was developed in a collaboration between 
Service d'A\'eronomie (Verri\`eres le Buisson, France) and the Finnish Meteorological Institute (Helsinki, Finland)
(Bertaux et al., 1995). Its main purpose is to study anisotropies in the solar wind mass flux and their effect on 
the interplanetary background (Bertaux et al., 1997). Thanks to the exceptional longevity of the SOHO mission, the 
data obtained by this instrument almost cover a complete 11-year solar activity cycle. 

The most recent analysis of the SWAN data pertaining to the determination of the ionizing fluxes from the Sun
was published by Qu\'emerais et al. (2006). In this study, the authors have determined the absolute value
of the ionization rate of hydrogen at 1 AU as a function of heliographic latitude and its variations with 
heliographic latitude. Their analysis covers the period from 1996 to 2005, i.e. almost a complete solar cycle. 
The Hydrogen ionization rate is related to the solar wind mass flux as shown in equation (2).
After correction for solar EUV photo-ionization and the variation of the charge exchange cross-section with
solar wind bulk velocity, the SWAN data give monthly averages of the solar wind mass flux at all latitudes 
between 1996 and 2005.

\subsection{Solar Wind Density: LASCO data}

In principle, the solar wind number density at 1 AU can be estimated from in situ measurements performed
by numerous spacecrafts proeminently restricted to the ecliptic plane.  However results from the Ulysses 
spacecraft have demonstrated that the density at higher latitudes can be very different from  that 
in the ecliptic (Phillips et al., 1995). 

We have therefore relied on the LASCO-C2 white light images to obtain an estimate of the solar wind density 
at all latitudes.  The LASCO-C2  polarized  radiance  images $pB$ are routinely generated from the  daily  
polarization  sequences consisting  a  set  of  three polarized  images taken  through the orange  filter  
(540-640 nm) and polarizers  oriented  at 60${}^o$,0${}^o$  and  -60${}^o$.
Qu\'emerais and Lamy (2002) have developed a method to invert these images and to produce two-dimensional 
electron density maps of the solar corona. 
Compared to our past studies (see also Lamy et al., 1997; LLebaria et al., 1999; Lamy et al., 2002), 
the maps used for this present studies have benefited from refined corrections which successfully remove 
the instrumental polarization.
The values at 6\rso\ from the center of the Sun have been extracted at all heliographic 
latitudes. We have then computed a monthly average from the daily maps to obtain values with the same 
sampling as the ionization rate derived from the SWAN data. 
We discuss below in which cases the density at 6\rso\ can be simply extrapolated at 1 AU.

\subsection{Solar Wind Terminal Velocity: IPS data}

As one can see from a quick perusing through the catalog of the omniweb database, various 
spacecraft have measured the solar wind velocity in the ecliptic plane, but only a few 
measurements are available at mid and high heliographic latitude.  
Here we have used the solar wind velocity estimates obtained from Interplanetary 
Scintillation data (IPS) (Kojima and Kakinuma, 1990). These velocity values are derived from the analysis of 
scintillations of  radio-emissions from compact radio sources (Kojima et al., 1998). Because the distribution of radio 
sources in the sky is not uniform, the sampling of these measurements is not either. These authors have found that
the solar wind speed is roughly constant beyond 0.1 AU from the sun which means that the acceleration occurs before.
Here we have averaged the velocity values beyond 0.1 AU from the sun as a function of the heliographic latitude from
which the wind originates. Those averages were made on a monthly basis.  It should be noted also that there is a 4 month 
data gap every year due to the weather conditions which prevent the ground-based radio observations.

\section{Results}

\subsection{Computation of H Ionization Rates}

The first step in our analysis is to check whether SWAN results are compatible with the LASCO density 
values, the IPS velocity estimates and the assumption that the IPS velocity is reached at the outer 
edge of LASCO-C2. To do that, we compute the charge exchange ionization rate at one AU $\beta_{exc}$ 
where the velocity is derived from the IPS data and the density is equal to the
value found from the LASCO images with a radial scaling coefficient. This coefficient expresses the relationship
between the density measured by LASCO at 6\rso\ and the density at 1 AU. For a radial expansion, the 
product $N \cdot V \cdot r^2$ is conserved. If the acceleration is finished, then the velocity is constant and
the density varies like $1/r^2$.

\begin{figure}
\noindent\rotatebox{90}{\includegraphics[height=15.0cm]{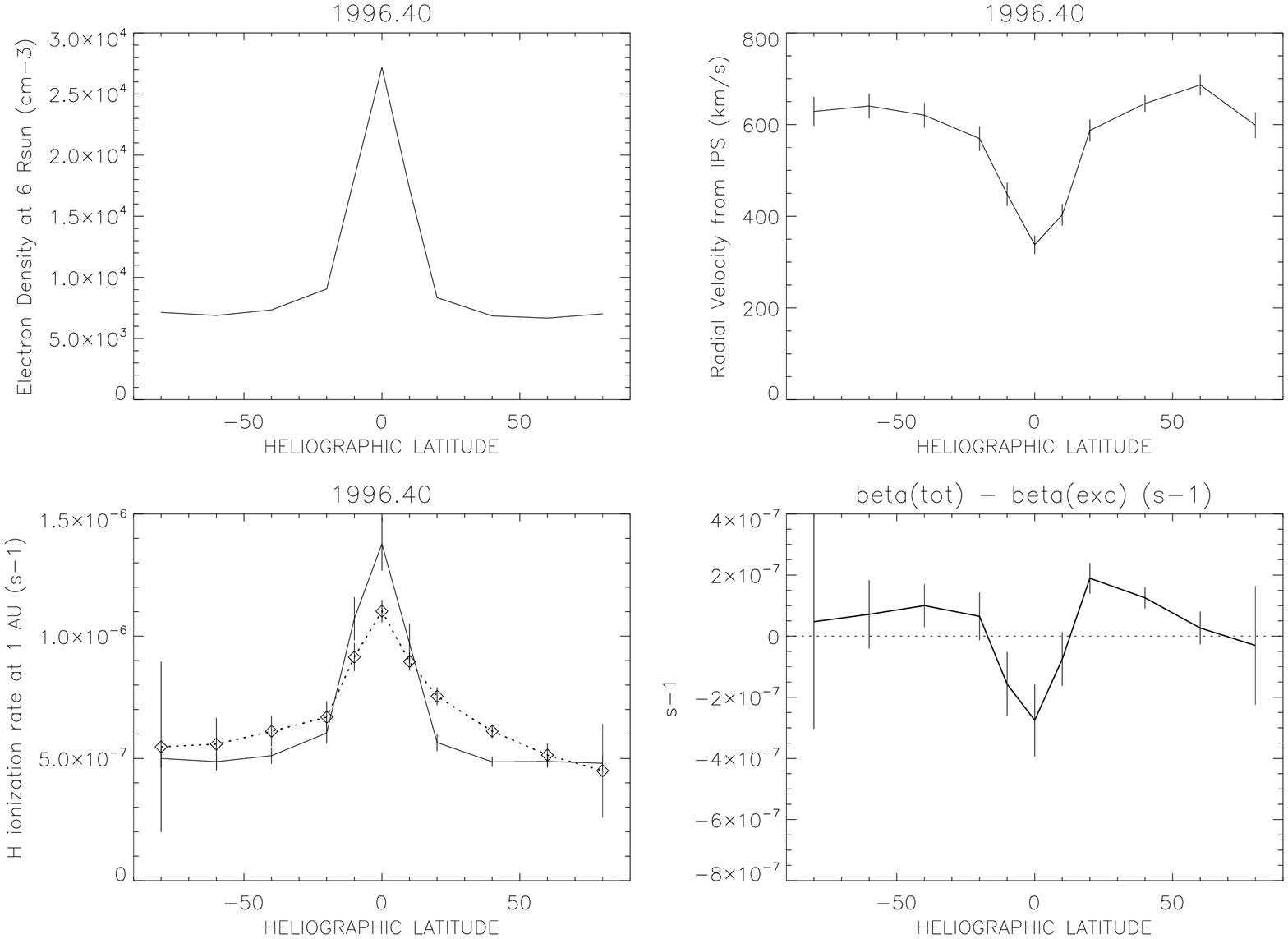}}
\caption{ Monthly averages of the data used in this analysis. These values were obtained in March and April 1996.
For all panels, the x axis shows the heliographic latitude. The top left panel shows the electron density at 6\rso\
obtained from the LASCO-C2 polarized white light images. The density is strongly enhanced around the equatorial belt.
The top right panel shows the average solar wind velocity as deduced from the interplanetary scintillation data. 
We clearly see the fast wind at high latitudes and the slow wind around the equator. The bottom left panel shows the
ionization rate of hydrogen at 1 AU. The diamonds joined by a dotted line shows the values derived from the SWAN data.
The solid line shows the charge exchange ionization found when using the densities (scaled at 1 AU) and the velocities
displayed in the top panels. The bottom right shows the difference between the SWAN values and the values obtained by
combining the LASCO-C2 and IPS data. Note the excess produced by the SWAN values at high latitudes, 
i.e., for the fast solar wind.}
\label{quadri96}
\end{figure}

Figure \ref{quadri96} shows the paramaters and calculations obtained in April-May 1996. The top left panel shows the
LASCO-C2 average density at 6\rso\ as a function of heliographic latitude. There is a strong enhancement in the 
equatorial band. The top-right panel shows the IPS mean velocity as a function of heliographic latitude for
the same period. The bottom-left panel shows the ionization rate of hydrogen at 1 AU. The diamonds linked by a dotted 
line correspond to the SWAN data. The solid line corresponds to the value derived from the scaled LASCO density and the
IPS velocity data.  The computed value represents only ionization through charge exchange. The bottom-right panel
shows the difference of the two curves shown in the bottom-left panel. In this plot, we see two different regimes.
In the equatorial band, the computed charge exchange rate is systematically larger than the SWAN total ionization rate.
This means that the velocity between 6\rso\  and 1 AU is not constant and that the radially scaled density is 
overestimated. This also implies that the velocity at 1 AU is larger than the velocity at 6\rso\  which means
that in the equatorial band, the solar wind has not completed its acceleration. Outside of the equatorial band,
the SWAN ionization rates are systematically larger than the computed charge exchange ionization rate. This is 
consistent with the fact that in this region the solar wind has reached its terminal velocity and that the density
simply scales like $1/r^2$ with solar distance. The difference between the two values can be attributed to the
sum of photo-ionization and electron impact ionization.

\begin{figure}
\noindent\rotatebox{90}{\includegraphics[height=15.0cm]{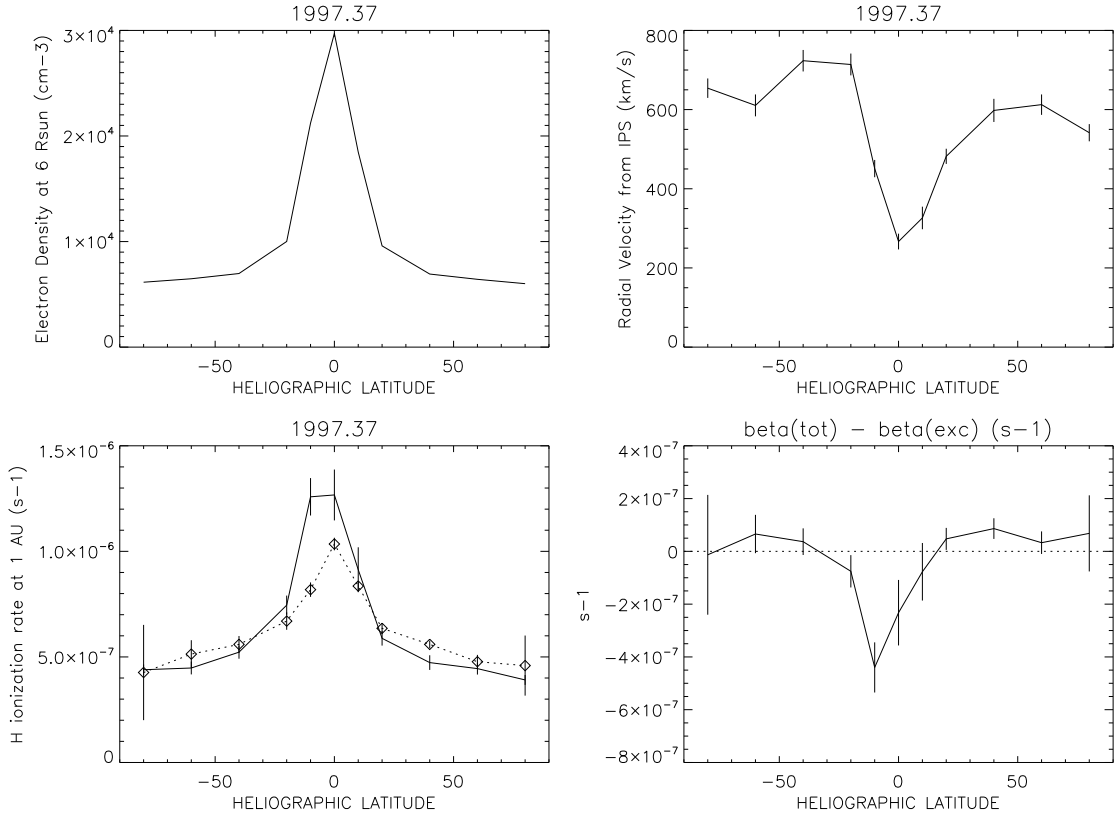}}
\caption{ Same as the previous figure but for data obtained in Spring 1997. Although the numerical values are not identical,
we find the same dichotomy between fast wind and slow wind. As shown in the bottom right panel, the flux deduced from the IPS 
values and and the LASCO-C2 data is too high along the equator. This means that the slow wind has not reached its 
terminal velocity at 6\rso\ .}
\label{quadri97}
\end{figure}

Figure \ref{quadri97} shows a very similar picture. The values shown here are for April-May 1997. Once again, the 
equatorial belt data suggest that the wind is not completely accelerated while at higher latitudes the ionization 
rate values agree well.

Figure \ref{diffbeta} shows the difference between the SWAN total ionization rate $\beta_{tot}$ and the
charge exchange ionization rate $\beta_{exc}$ computed from LASCO and IPS data as a function of the IPS solar 
wind velocity. All monthly values for different heliographic latitudes have been used. The data are averaged 
over intervals of 10 degrees in latitude. Two patterns appear clearly. For velocities larger than 500 km/s, the quantity 
$\beta_{tot} - \beta_{exc}$  is positive. Its mean value is equal $(5.3 \pm 2.3) \times 10^{-8}$ s$^{-1}$.
This is consistent with the fact that for the fast wind conditions (here faster than 500 km/s), the acceleration
is completed at 6\rso\ and that the solar wind density at 1 AU is  correctly derived from the LASCO value 
scaled by $1/r^2$. In that case, the difference between total ionization rate and charge exchange ionization 
rate is equal to the sum of hydrogen photo-ionization rate and electron impact ionization rate. The value
of $(5.3 \pm 2.3) \times 10^{-8}$ s$^{-1}$ found here is compatible with previously estimated values. For example
hydrogen photo-ionization rate estimates for solar minimum conditions
range between $5 \times 10^{-8}$ s$^{-1}$ (Rucinski and Fahr, 1991) to $10^{-7}$ s$^{-1}$ (Ogawa et al., 1995).
We estimated electron impact ionization rates for hydrogen using the WIND/WAVES data (Issautier et al., 2005) and
found values around $10^{-8}$ s$^{-1}$ close to solar minimum.

When the IPS velocity value is lower than 500 km/s, we find that the quantity $\beta_{tot} - \beta_{exc}$ is
negative. As mentioned above, the flux conservation from $r_i = 6$\rso\ to 1 AU also contains a ratio 
$V(r_i)/V(1AU)$. This suggests that for slow solar wind conditions, the solar wind has not yet reached its terminal
velocity, and that the ratio $V(r_i)/V(1AU)$ is smaller than one, assuming a radial outflow as defined in equation (3).

Here we find a clear dichotomy between fast solar wind regions at high latitude and slow solar wind regions near the
equatorial region.

\subsection{Velocity Profiles in the Corona}

\begin{figure}
\noindent\includegraphics[height=12.0cm]{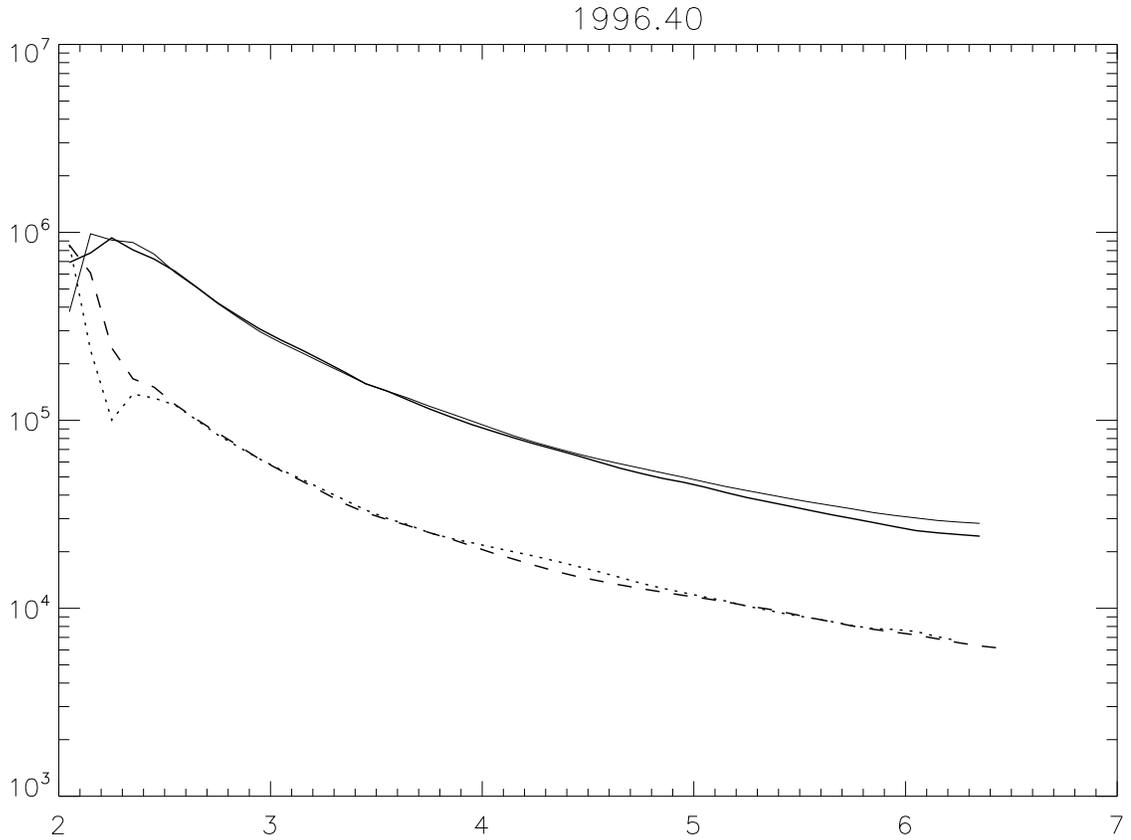}
\caption{ Density profiles for Spring 1996 deduced from the LASCO-C2 polarized white light images. The values correspond 
to one-month averages. The top solid lines give the values for the East and West limbs. The two limbs are very similar 
which is normal for an average of one month of data. The dashed line shows the north pole density profile and the dotted 
line shows the south pole profile. There's almost a factor of 10 between polar and equatorial values.}
\label{densite}
\end{figure}

Figure \ref{densite} shows the LASCO-C2 electronic density radial profiles in spring 1996 averaged
over the same period as the SWAN data. This results in very regular curves where most 
transient effects like CME's have been smoothed out. The two solid lines show the East limb and West limb density
profiles. The dashed line shows the North pole density profile and the dotted line shows the south pole density profile.
Within 2.5 \rso\ of the Sun, the data are contaminated by the instrument scatterred light and the uncertainty is large. 
The profiles can only be used with confidence between 2.5 \rso\ and 6 \rso\ .  
We see that the pole profiles have densities which are 10 times lower than the equatorial values.

\begin{figure}
\noindent\includegraphics[height=12.0cm]{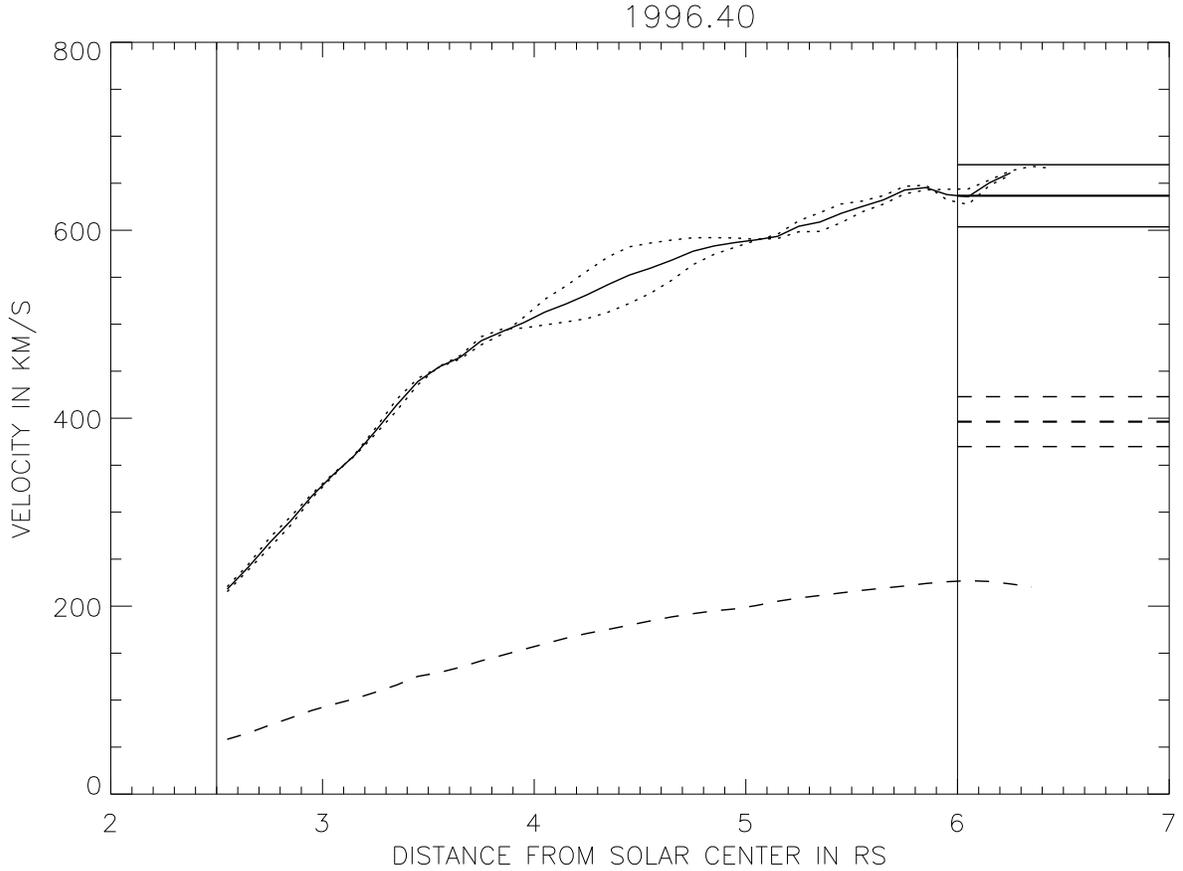}
\caption{ Velocity profiles for Spring 1996 deduced from the combination of SWAN, LASCO and IPS data.  The solid line 
represents the average of the north and south profiles while the dotted lines show the individual north and south profiles. 
We see an inflexion around 3.5\rso\ which suggests a change in the acceleration at this altitude. The dashed line 
shows the profile derived for the equatorial solar wind. This profile was derived by assuming a radial expansion and may 
have to be corrected for non-radial  effects. After 6\rso\ , the lines show the IPS velocity values. For both polar 
profiles, we see that the profiles reach the terminal value around 6\rso\ , whereas at the equator the velocity at 
6\rso\ is only half of its terminal value.}
\label{vitesse}
\end{figure}

Figure \ref{vitesse} displays the velocity profiles derived from the densities in Figure \ref{densite} using equation 
(4). The flux scaling factor $\frac{\beta_{exc}}{\sigma_{exc}(V_o)}$ is derived from the SWAN data of the same date.
The solid line shows the radial dependence of the solar wind as a function of distance for the average of North and 
South poles. On the right hand we have plotted the corresponding IPS value around 600 km/s. The dotted line shows 
the individual profiles derived for the north and south poles. They are very similar to the average profile. 
From these profiles, we see that the solar wind goes from about 
200 km/s at 2.5\rso\ up to $\sim$ 600 km/s at 4.5\rso\ . After a distance of 5\rso\ ,  the increase of solar wind 
velocity is less pronounced and acceleration is essentially finished at 6\rso\ .

The dashed line shows the profile derived using the average of the two equatorial density profiles shown in figure 
\ref{densite}. Once again the flux scaling ratio is derived from the SWAN equatorial data for the same period.
The corresponding IPS velocity value is shown on the right hand part of the plot and has a value close to 400 km/s.
We see that for the slow wind, the mean velocity at 6\rso\ is only 200 km/s, i.e. half of the terminal velocity.
This result is limited by the assumption of radial expansion which is implicit in equation (4). 
Although this assumption is quite justified for a polar coronal hole above 5\rso\ or 6\rso\ , 
this may not be the case for the equatorial slow wind. 
In that case a correction term should be applied to take into account the non-radial part of the slow wind expansion. 
Furthermore, this profile concerns an average of many transient phenomena which may have had very different individual 
velocity profiles.

\section{Comparison with other works}

The present work  provides time-averaged velocity profiles at all latitudes. Coronal 
flow velocities have been obtained in specific regions and at specific times in different ways. 
The outflow velocity has been inferred from analyses of the Doppler dimming of the intensities of the OVI 1032\AA , 
1037 \AA\ ,  HI L$\alpha$ 1216\AA\ lines, using the UVCS on board SOHO. In particular, Strachan et al. (2000) 
derived the latitudinal dependence of O5+ outflow velocities between 1.75 \rso\ and 2.75 \rso\ . 
Gabriel et al. (2005) derive an outflow velocity of 150-200 km/ at 2.5 $R_s$ in polar plumes.
Kohl et al. (1998), Cranmer et al. (1999) and Antonucci et al. (2000) show that in a polar hole, the fast wind speed 
reaches 300-400  km/s near 3\rso\ . On the other hand, Raymond et al. use UVCS observations of a sun-grazing 
comet at 6.8\rso\ and derive an outflow velocity smaller than 640 km/s in an high speed wind region. 
They argued that only 50\% to 75\% of the 1 AU proton energy is reached at 6.8\rso\ and, that an additional energy
deposition is at work beyond this distance.
 
Above streamers, Strachan et al. (2002) derived a null velocity up to 3.5\rso\ and an abrupt acceleration to 90 
km/s at 5\rso\ along the streamer axis, while along the streamer legs the velocity has already reached 100 km/s at 
2.33\rso\ . Based on expanding features, Sheeley et al. (1997) estimated the acceleration of the slow wind 
above streamers and found it compatible with a Parker-type isothermal expansion  with values
ranging from 150 to 300 km/s between 5\rso\ and 25\rso\ . 

In the open field/fast wind regions, the geometry is extremely simple and the comparison between previous 
localised measurements and the monthly-averaged SWAN/LASCO/IPS results can be directly performed. Indeed, there 
is a quite good agreement with SUMER and UVCS results between 2\rso\ and 3\rso\ . The agreement suggests that the 
fast wind acceleration is rather homogeneous and constant with time. Note that the flow speed we derived is 
however above the maximum speed at 6.8\rso\ deduced from the Sun-grazing comet.

For the slow wind the comparison is not so easy because it originates in and around closed-field regions 
with a more complicated geometry. It has been shown by Strachan et al. (2002) that the profiles vary strongly 
from the streamer axis to the legs.  Since the streamers vary in angle, latitude, and shape more work is 
needed to interpret the average profile. 

\section{Conclusion}

In this article,
we have shown how we can retrieve velocity profiles in the corona between 2.5\rso\  and 6\rso\ .

This analysis has been applied to data from 1996 and 1997. We have used monthly averages because the SWAN inversion technique
is based on data covering at least one solar rotation.
We have found that during solar minimum, there is a strong difference between the velocity profiles derived for the polar 
and the equatorial regions of the Sun. The polar regions show a fast solar wind with lower density at 1 AU while  the equatorial
regions are the source of the slow solar wind with increased density. This features was already well known from analysis
of the Interplanetary Scintillation data (Kojima and Kakinuma, 1990) or the Ulysses fast latitudinal scan of 1994 
(Phillips et al., 1995).

Using the LASCO density values derived at 6\rso\ , we have been able to show that the fast wind has a radial expansion with
constant velocity beyond 6\rso\ . This confirms the idea that the fast solar wind is accelerated at low altitudes in the corona
and that it has already reached its terminal velocity at 6\rso\ . This also means that the density ratio between 6\rso\ and
1 AU is simply scaled by the ratio of the square of the distance to the Sun (equation 3, where $V(R) = V_o$ and R = 6 \rso\ ). 
On a side note, we can also mention that in this case, a combination of the SWAN flux data with the LASCO density values 
can be used to derive the velocity value at 1 AU without the need of the information given by the IPS analysis. 

The slow wind originating from equatorial regions of the Sun shows a different behaviour. First, we have seen that the density
ratio between 6\rso\ and 1 AU is not simply equal to the ratio of the square of the distance to the sun. This is shown
by computing the flux using the density and the IPS velocity and comparing to the SWAN value. We see that the SWAN value is 
always smaller which means that the velocity ratio between the outer edge of LASCO-C2 and 1 AU,  $V(6R_{\odot})/V_o$,  
is smaller than one. 

\begin{figure}
\noindent\rotatebox{90}{\includegraphics[height=15.0cm]{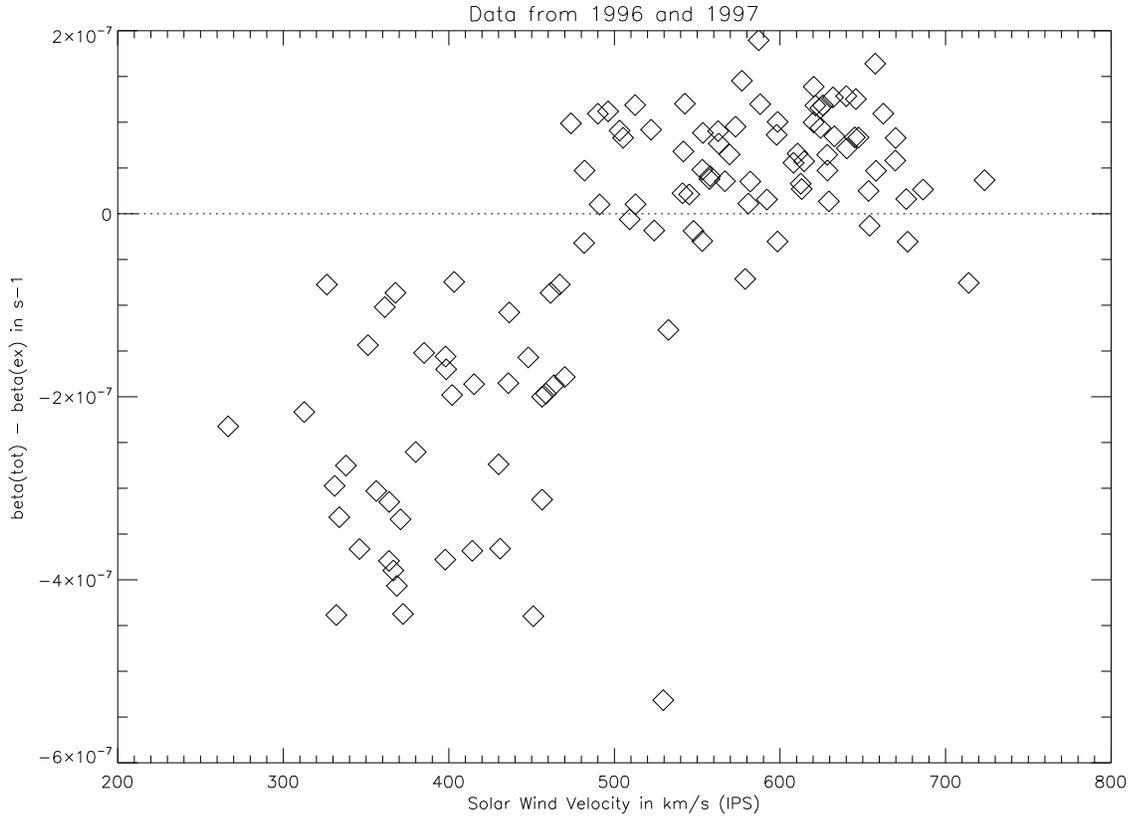}}
\caption{ Difference between the SWAN total ionization rate at 1 AU and the charge exchange rate deduced from LASCO density
 and IPS velocity data. The x-axis shows the velocity value from the IPS data. In this plot, we have used all available 
data from 1996 and 1997, averaged monthly. The latitude bins are 10 degrees wide. Here we see the two types of wind. 
The fast wind ($\geq$ 500 km/s) where the difference is positive and the slow wind where it is negative. The average of
the fast wind values give an estimate for the sum of photo-ionization and electron impact ionization rates.}
\label{diffbeta}
\end{figure}

As shown in Figure \ref{diffbeta}, the threshold between fast and slow winds is $\sim$  500 km/s. Below that threshold, the
wind has not reached its terminal velocity at 6\rso\ while it has reached its terminal velocity at 6\rso\ for values larger 
than 500 km/s. The velocity profile for the slow wind may be different from the one we have produced if the expansion factor
is not purely radial. This may be true at lower altitudes in the corona.

Finally, our velocity profiles provide a test of the models of solar wind acceleration 
to reveal which mechanisms are really effective.  
In future developments of this work, we will extend this study to all periods of solar activity. 
The corresponding SWAN data used to derive the solar wind flux at 1 AU have already been analysed by 
Qu\'emerais et al. (2006).
This will allow us to study the variation of the acceleration profile with solar activity. 

\begin{flushleft}

\noindent
{\em Acknowledgements}

SOHO is a mission of international cooperation between ESA and NASA.
SWAN and LASCO-C2 activities in France are funded by CNES with support from CNRS.
SWAN activities in Finland are funded by TEKES and the Finnish Meteorological Institute.

\end{flushleft}


\clearpage

\clearpage

\end{document}